\documentclass{article}
\usepackage{amsmath}

\input{tcilatex}

\begin{document}

\title{Theoretical investigations on the Y(4260) being an hybrid meson.}
\author{F. Iddir\thanks{%
E-mail: $faridaghis@hotmail.com$} and L. Semlala\thanks{%
E-mail: $l\_semlala@yahoo.fr$} \\
Laboratoire de Physique Th\'{e}orique, \\
Universit\'{e} d'Oran Es-S\'{e}nia, 31100, ALGERIA}
\maketitle

\begin{abstract}
The new recently experiments at the B-factories yield a renewed interest in
the charm and charmonium spectroscopy. New intriguing states have been
observed, which appear to be non-conventional mesons, such as $X\left(
3862\right) \ $and $Y\left( 4260\right) $ and request more theoretical
investigations.

We will explore the possibility for the $Y\left( 4260\right) $ to be a $c%
\overline{c}g$ hybrid meson. Using the quark-gluon constituent model, we
exclude its existence as QE-hybrid meson, and as mixing of $c\overline{c}g$
hybrid with conventional $c\overline{c}$ meson. We find the only
interpretation as GE-hybrid, decaying in $D_{1}\overline{D}$ channels. Then
more experimental studies are needed to confirm the existence of this
resonance and to give its properties.
\end{abstract}

\section{Introduction}

In the last few years, an intense activity has been done at the B-factories,
and has seen important progress of charm and charmonium spectroscopy. New
charm mesons and baryons have been observed, and mainly charmonium-like
mesons that need particular attention, since they do not classify in the
standard hadronic spectroscopy (by the naive quark model). The more
interesting observed new states are the $X\left( 3872\right) $\ and the $%
Y\left( 4260\right) $.

$\bullet $\underline{The $X\left( 3872\right) $} was first discovered by
BELLE$^{\left[ 1\right] }$ in $B^{\pm }\longrightarrow K^{\pm }X\left(
3872\right) $, decaying in $\pi ^{+}\pi ^{-}J/\psi $, and confirmed by BABAR$%
^{\left[ 2\right] }$, CDF$^{\left[ 3\right] }$ and DO$^{\left[ 4\right] }$
collaborations. The mass of the $X\left( 3872\right) $ is within errors of
the $D^{0}\overline{D^{\ast 0}}$ threshold $\left( 3871.3\pm 1.0\text{ }%
MeV\right) ^{\left[ 5\right] }$. Other decay modes have been observed:

\textit{i}) in $\gamma J/\psi ^{\left[ 6\right] }$ with a branching ration
of $\left( 19\pm 7\right) \%$ relative to the $X\left( 3872\right)
\longrightarrow $ $\pi ^{+}\pi ^{-}J/\psi $ mode,

\textit{ii}) in $\overline{D^{0}}D^{0}\pi ^{0}$ $^{\left[ 7\right] }$; this
mode being dominant to have a branching ratio $\left( 9.7\pm 3.4\right) $
times higher than the $\pi ^{+}\pi ^{-}J/\psi $ mode.

Recent studies from BELLE, that combine angular and kinematic properties of
the $\pi ^{+}\pi ^{-}$\ mass strongly favor the assignment of $J^{PC}=1^{++}$
$^{\left[ 6\right] }$, which is in agreement with expectations of models
interpreting the $X\left( 3872\right) $\ as a molecule-like $D^{0}\overline{%
D^{\ast 0}}$\ bound state$^{\left[ 7\right] }$. Some other interpretations
have been proposed for the $X\left( 3872\right) $. Hence, the $X\left(
3872\right) $\ appears not to be a simple\ meson state.

$\bullet $\underline{The $Y\left( 4260\right) $}$.$

The second interesting charmonium-like state is a new structure observed by
BABAR Collaboration$^{\left[ 10\right] }$ in ISR (Initial State Radiation)
events at $4259\pm 8_{-6}^{+2}$ $MeV$ with a width of $\Gamma =88\pm
23_{-4}^{+6}$ $MeV$, decaying in $\pi ^{+}\pi ^{-}J/\psi $, and having $%
J^{PC}=1^{--}$. The $Y$ state has been confirmed by CLEO$^{\left[ 11\right]
} $ in $\pi ^{+}\pi ^{-}J/\psi $ and $\pi ^{0}\pi ^{0}J/\psi $, which
implies that it has Isospin $I=0$.

Investigations have been done to observe other decay modes, like $D\overline{%
D\text{ }}$ $^{\left[ 12\right] }$, $p\overline{p}^{\left[ 13\right] }$, $%
\Phi \pi ^{+}\pi ^{-\left[ 14\right] }$, without success, and limits for the
decay rates are given; the upper limit for the $D\overline{D\text{ }}$\ mode
is 7.6 times relative to the $Y\longrightarrow $ $\pi ^{+}\pi ^{-}J/\psi $
mode, which is much smaller than \textit{BR}$\left[ \Psi \left( 3770\right)
\longrightarrow D\overline{D\text{ }}\right] \approx $500$\times $\textit{BR}%
$\left[ \Psi \left( 3770\right) \longrightarrow \pi ^{+}\pi ^{-}J/\psi %
\right] $ $^{\left[ 15\right] }$; then the $Y\left( 4260\right) $ cannot be
a radially excited state such as $\Psi \left( 4S\right) $. Furthermore, it
should be noted that neither of $\Psi \left( 4040\right) $, $\Psi \left(
4160\right) $, $\Psi \left( 4415\right) $, has been observed decaying in $%
\pi ^{+}\pi ^{-}J/\psi $ mode.

Several explanations have been proposed for the $Y\left( 4260\right) $, the
most favored being the $c\overline{c}g$\ hybrid meson structure$^{\left[
16,17\right] }$. We will in this work explore this hypothesis.

In section 2, we present the model used to represent the hybrid meson, as
the quark-gluon constituent model, with the distinction between the two
different modes (GE and QE).

In section 3, we give the selection rules for the decay of an hybrid $1^{--}$
meson in two mesons and we compute the decay widths.

In section 4, we explore the possibility for the $Y\left( 4260\right) $ to
be a mixing of an hybrid with a conventional charmed meson, as argued in $%
\left[ 16\right] $. We conclude in section 5.

\section{The hybrid state from the constituent model}

In the framework of the strong-coupling regime, we can not yet be able to
perform systematical calculations directly from the QCD Lagrangian, the
process need alternative models, like Flux-Tube Model, Bag Model, QCD Sum
Rules, Lattice QCD, ... or phenomenological potential models.

The quark model based on the instantaneous potential interaction between the
hadron's constituents is motivated by the experiment. Indeed, practically
all our knowledge on hadron structure comes from phenomenological models,
the constituent quark model particularly continues to have great success. We
will use it to study the $J^{PC}=1^{--}\ Y\left( 4260\right) $.

In a recent work$^{\left[ 21\right] }$, we used a phenomenological potential
which reproduces the QCD characteristics (asymptotic freedom and
confinement), its expression having the mathematical $\left( \text{%
Coulomb+linear}\right) $ form. Relativistic corrections both in the kinetic
and interaction parts of the Hamiltonian are considered.(taking into account
spin-spin, spin-orbit and tensor interactions).

An important ingredient of the model is the mass of the constituent gluon $%
m_{g}$. There is evidence for massive like dispersion relation for the
gluon, with mass ranging from $700\sim 1000$ $MeV$, both from lattice and
Schwinger-Dyson equations. As for quarks, this represents a dynamical mass
and is $\exp $ected to add $0.7\sim 1.0$ $GeV$ to the corresponding
quarkonia.

We represent a state of the hybrid meson by the following quantum numbers:

$l_{g}$ is the relative orbital momentum of the gluon in the $q\overline{q}$%
\ center of mass;

$l_{q\overline{q}}$ is the relative orbital momentum between $q$ and $%
\overline{q}$;

$S_{q\overline{q}}$ is the total quark spin;

$J_{g}$ is the total gluon angular momentum;

$L=l_{q\overline{q}}+J_{g}$.

Parity and Charge Conjugation of the hybrid meson verify:

\begin{center}
\begin{equation}
P=\left( -\right) ^{l_{q\overline{q}}+l_{g}},\text{ }C=\left( -\right) ^{l_{q%
\overline{q}}+S_{q\overline{q}}+1}  \tag{1}
\end{equation}
\end{center}

For the lowest $J^{PC}=1^{--\text{ }}$ hybrid states, equation $\left(
1\right) $ implies $l_{q\overline{q}}=S_{q\overline{q}}$ \ and $l_{q%
\overline{q}}+l_{g}$ odd, which gives two possibilities (see Table 1):

\begin{center}
\begin{equation}
l_{q\overline{q}}=S_{q\overline{q}}=0\text{ , }l_{g}=1  \tag{2}
\end{equation}
\end{center}

and

\begin{center}
\begin{equation}
l_{q\overline{q}}=S_{q\overline{q}}=1\text{ , }l_{g}=0  \tag{3}
\end{equation}
\end{center}

We refer to the $\left( 2\right) $ as the gluon-excited mode $\left( \text{GE%
}\right) $, and to the $\left( 3\right) $ as the quark-excited mode $\left( 
\text{QE}\right) $.

\begin{center}
\ $\underset{\text{\textit{Table\ 1:\ The\ lowest\ }}J^{PC}=1^{--\text{ }}\ 
\text{\textit{hybrid\ states\ with\ their\ quantum\ numbers}.}}{_{ 
\begin{tabular}{|c|c|c|c|c|c|c|c|}
\hline
$P$ & $C$ & $J$ & $l_{g}$ & $l_{q\overline{q}}$ & $J_{g}$ & $S_{q\overline{q}%
}$ & $L$ \\ \hline
$-$ & $-$ & $1$ & $0$ & $1$ & $1$ & $1$ & $0$ \\ \hline
$-$ & $-$ & $1$ & $0$ & $1$ & $1$ & $1$ & $1$ \\ \hline
$-$ & $-$ & $1$ & $0$ & $1$ & $1$ & $1$ & $2$ \\ \hline
$-$ & $-$ & $1$ & $1$ & $0$ & $1$ & $0$ & $1$ \\ \hline
\end{tabular}
}}$
\end{center}

We found in [21] a significant mixing between the two QE and GE-modes, and
we predict a $1^{--}c\overline{c}g$ mass around $4.27$ $GeV$ (see Table 2):

\begin{center}
$\underset{\text{\textit{Table 2: Mixed QE-GE hybrid mesons masses with spin
corrections (in GeV)}}}{^{ 
\begin{tabular}[t]{|c|c|c|c|c|}
\hline
& $1^{--}$ & $0^{--}$ & $0^{-+}$ & $1^{-+}$ \\ \hline
$n\overline{n}g$ & 1.40 & 1.58 & 1.73 & 1.93 \\ \hline
$s\overline{s}g$ & 1.66 & 1.87 & 2.02 & 2.21 \\ \hline
$c\overline{c}g$ & 4.27 & 4.35 & 4.38 & 4.48 \\ \hline
$b\overline{b}g$ & 10.50 & 10.66 & 10.68 & 10.80 \\ \hline
\end{tabular}
}}$
\end{center}

The order of magnitude being in agreement with the masses obtained by the
other models$^{\left[ 22\right] }$.

\section{The decay of the hybrid meson}

\subsection{The decay process}

In the model, at the leading order, the decay of the hybrid meson into two
standard mesons may occur through two diagrams, the gluon annihilating into
a $q\overline{q}$ pair (namely $q=u,d$ ) (Fig.1, Fig.2).

The process of the Fig.1 dominates, the second (disconnected diagram) being
suppressed by the OZI rule.

Fig.3 represents the process able to product directly $\pi ^{+}\pi
^{-}J/\psi $, the mechanism being strongly suppressed.

This representation of the decay could not explain the experimental data, no
signal of two-mesons decay mode being detected, especially as two-body decay
usually dominates three-body decay.

On the other hand, a $c\overline{c}g$ $1^{--}$\ hybrid meson in the mass
range of $\sim 4.$ $GeV$ can couple to $D_{1}\left( 2420\right) \overline{D}$
and $D_{1}\left( 2420\right) ^{\pm }D^{\mp }$ (through the process of Fig.1)
and a $\pi ^{+}\pi ^{-}J/\psi $ signal may be produced by re-scattering
effects, but the dominant final state should be in $DD^{\ast }\pi $. This
hypothesis suggests more experimental data to confirm the mass of the $Y$
(the threshold of a $D_{1}\left( 2420\right) \overline{D}$ channel needs a
mass lightly above) and to search for other decay channels.

\subsection{The decay widths}

The decay of an hybrid state A into two mesons B and C is represented by the
matrix element of the Hamiltonian annihilating a gluon and creating a quark
pair:

\begin{equation}
\langle BC\left\vert H\right\vert A\rangle =gf(A,B,C)\left( 2\pi \right)
^{3}\delta _{3}\left( p_{A}-p_{B}-p_{C}\right) ;  \tag{4}
\end{equation}

where $f(A,B,C)$ representing the decay amplitude by:

\begin{eqnarray}
f(A,B,C) &=&\sum_{\left( m\right) ,\left( \mu \right) }\Phi \Omega X\left(
\mu _{q\overline{q}},\mu _{g};\mu _{B},\mu _{C}\right) I\left( m_{q\overline{%
q}},m_{g};m_{B},m_{C},m\right)  \TCItag{5} \\
&&\times \left\langle l_{g}m_{g}1\mu _{g}\right\vert J_{g}M_{g}\rangle
\langle l_{q\overline{q}}m_{q\overline{q}}J_{g}M_{g}\left\vert Lm^{\prime
}\right\rangle \langle Lm^{\prime }S_{q\overline{q}}\mu _{q\overline{q}%
}\left\vert JM\right\rangle  \notag \\
&&\times \langle l_{B}m_{B}S_{B}\mu _{B}\left\vert J_{B}M_{B}\right\rangle
\langle l_{C}m_{C}S_{C}\mu _{C}\left\vert J_{C}M_{C}\right\rangle .  \notag
\end{eqnarray}

where $\Phi ,$ $\Omega ,$ $X$ and $I$ are the flavor, color, spin and
spatial overlaps. $\Omega $ is given by:

\begin{equation}
\Omega =\frac{1}{24}\sum_{a}tr\left( \lambda ^{a}\right) ^{2}=\frac{2}{3}. 
\tag{6}
\end{equation}

From:

\begin{equation}
\chi _{\mu _{1}}^{+}\sigma ^{\lambda }\chi _{\mu _{2}}=\sqrt{3}\langle \frac{%
1}{2}\mu _{2}1\lambda \left\vert \frac{1}{2}\mu _{1}\right\rangle ,  \tag{7}
\end{equation}

we obtain the spin overlap:

\begin{eqnarray}
X\left( \mu _{q\overline{q}},\mu _{g};\mu _{B},\mu _{C}\right) &=&\sum_{S}%
\sqrt{2}\left[ 
\begin{array}{ccc}
1/2 & 1/2 & S_{B} \\ 
1/2 & 1/2 & S_{C} \\ 
S_{q\overline{q}} & 1 & S%
\end{array}
\right]  \TCItag{8} \\
&&\times \langle S_{q\overline{q}}\mu _{q\overline{q}}1\mu _{g}\left\vert
S\left( \mu _{q\overline{q}}+\mu _{g}\right) \right\rangle \langle S_{B}\mu
_{B}S_{C}\mu _{C}\left\vert S\left( \mu _{B}+\mu _{C}\right) \right\rangle ;
\notag
\end{eqnarray}

where

\begin{equation}
\left[ 
\begin{array}{ccc}
1/2 & 1/2 & S_{B} \\ 
1/2 & 1/2 & S_{C} \\ 
S_{q\overline{q}} & 1 & S%
\end{array}
\right] =\sqrt{3\left( 2S_{B}+1\right) \left( 2S_{C}+1\right) \left( 2S_{q%
\overline{q}}+1\right) }\left\{ 
\begin{array}{ccc}
1/2 & 1/2 & S_{B} \\ 
1/2 & 1/2 & S_{C} \\ 
S_{q\overline{q}} & 1 & S%
\end{array}
\right\} .  \tag{9}
\end{equation}

The spatial overlap is given by:

\begin{eqnarray}
I\left( m_{q\overline{q}},m_{g};m_{B},m_{C},m\right) &=&\iint \frac{d%
\overrightarrow{p}d\overrightarrow{k}}{\left( 2\pi \right) ^{6}\sqrt{2\omega 
}}\Psi _{q\overline{q}g}^{l_{q\overline{q}}m_{q\overline{q}%
}l_{g}m_{g}}\left( \overrightarrow{P}_{B}-\overrightarrow{p},\overrightarrow{%
k}\right)  \TCItag{10} \\
&&\times \Psi _{q_{i}\overline{q}}^{l_{B}m_{B}\ \ast }\left( \overrightarrow{%
p}_{1}\right) \Psi _{q\overline{q}_{i}}^{l_{C}m_{C}\ \ast }\left( 
\overrightarrow{p}_{2}\right) Y_{l}^{m\ \ast }\left( \Omega _{B}\right)
d\Omega _{B},  \notag
\end{eqnarray}

where:

\begin{equation}
\overrightarrow{p}_{1}=\frac{m_{\overline{q}_{i}}}{m_{q}+m_{\overline{q}_{i}}%
}\overrightarrow{P}_{B}-\overrightarrow{p}-\frac{\overrightarrow{k}}{2} 
\tag{11}
\end{equation}

\begin{equation}
\overrightarrow{p}_{2}=-\frac{m_{q_{i}}}{m_{\overline{q}}+m_{q_{i}}}%
\overrightarrow{P}_{B}+\overrightarrow{p}-\frac{\overrightarrow{k}}{2}. 
\tag{12}
\end{equation}

$l,m\ \ $label the orbital momentum between the two final mesons.

Finally,

\begin{equation}
\Phi =\left[ 
\begin{array}{ccc}
i_{1} & i_{3} & I_{B} \\ 
i_{2} & i_{4} & I_{C} \\ 
S_{q\overline{q}} & 1 & I_{A}%
\end{array}
\right] \eta \epsilon ;  \tag{13}
\end{equation}

where I's (i's) label the hadron (quark) isospins, $\eta =1$ if the gluon
goes into strange quarks and $\eta =\sqrt{2}$ if it goes into non strange
ones. $\epsilon $ is the number of diagrams contributing to the decay.
Indeed one can check that since P and C are conserved, two diagrams
contribute with the same sign and magnitude for allowed decays while they
cancel for forbidden ones. In the case of two identical final particles, $%
\epsilon =\sqrt{2}$.

The partial width is then given by:

\begin{equation}
\Gamma \left( A\rightarrow BC\right) =4\alpha _{s}\left\vert f\left(
A,B,C\right) \right\vert ^{2}\frac{P_{B}E_{B}E_{C}}{M_{A}};  \tag{14}
\end{equation}

with

\begin{equation}
P_{B}^{2}=\frac{\left[ M_{A}^{2}-\left( m_{B}+m_{C}\right) ^{2}\right] \left[
M_{A}^{2}-\left( m_{B}-m_{C}\right) ^{2}\right] }{4M_{A}^{2}};  \tag{15}
\end{equation}

\begin{eqnarray}
E_{B} &=&\sqrt{P_{B}^{2}+m_{B}^{2}};  \TCItag{16} \\
E_{B} &=&\sqrt{P_{B}^{2}+m_{C}^{2}}.  \notag
\end{eqnarray}

\subsection{Selection rules for the decay}

Computing the integral $\left( 10\right) $ in the two cases, for the decay
respectively into two S-Wave mesons and into (S-wave + P-wave) mesons, leads
to two selection rules.

Using the Harmonic oscillator potential, we find:

\begin{eqnarray}
I\left( m_{q\overline{q}},0;0,0,m\right) &=&2^{4}\sqrt{\frac{\pi }{3\omega }}%
\frac{R_{q\overline{q}}^{3/2+l_{q\overline{q}}}R_{g}^{3/2+l_{g}}R_{B}^{5}}{%
\left( R_{g}^{2}+R_{B}^{2}/2\right) ^{3/2}\left( R_{q\overline{q}%
}^{2}+2R_{B}^{2}\right) ^{5/2}}\frac{2m_{q}}{m_{q}+m_{q_{i}}}P_{B}  \notag \\
&&\times \exp -\frac{P_{B}^{2}}{2}\left[ R_{q\overline{q}}^{2}+\frac{2m_{%
\overline{q_{i}}}^{2}R_{B}^{2}}{\left( m_{q}+m_{\overline{q_{i}}}\right) ^{2}%
}-\frac{\left[ 2m_{q_{i}}R_{B}^{2}+\left( m_{q}+m_{\overline{q_{i}}}\right)
R_{q\overline{q}}^{2}\right] ^{2}}{\left( R_{q\overline{q}%
}^{2}+2R_{B}^{2}\right) \left( m_{q}+m_{\overline{q_{i}}}\right) ^{2}}\right]
\notag \\
&&\delta _{l_{g},0}\delta _{l_{q\overline{q}},l}\delta _{m_{q\overline{q}},m}
\TCItag{17}
\end{eqnarray}

and

\begin{eqnarray}
I\left( 0,m_{g};m_{B},0,0\right) &=&-\sqrt{\frac{\pi }{2\omega }}\frac{R_{q%
\overline{q}}^{3/2}R_{g}^{5/2}R_{B}^{4}}{\left(
R_{g}^{2}/2+R_{B}^{2}/4\right) ^{5/2}\left( R_{q\overline{q}%
}^{2}/2+R_{B}^{2}\right) ^{3/2}}  \notag \\
&&\times \exp \left\{ -\frac{P_{B}^{2}}{2}\left[ R_{q\overline{q}}^{2}+\frac{%
2m_{\overline{q_{i}}}^{2}R_{B}^{2}}{\left( m_{q}+m_{\overline{q_{i}}}\right)
^{2}}-\frac{\left[ 2m_{q_{i}}R_{B}^{2}+\left( m_{q}+m_{\overline{q_{i}}%
}\right) R_{q\overline{q}}^{2}\right] ^{2}}{\left( R_{q\overline{q}%
}^{2}+2R_{B}^{2}\right) \left( m_{q}+m_{\overline{q_{i}}}\right) ^{2}}\right]
\right\}  \notag \\
&&\delta _{l_{g},1}\delta _{l_{q\overline{q}},0}\delta _{m_{q\overline{q}%
},0}\delta _{m_{g}m_{B}}  \TCItag{18}
\end{eqnarray}

$q_{i}\overline{q_{i}}$ is the quark created pair.

Equation $\left( 17\right) $ corresponds to the QE-hybrid, which is the only
mode allowed to decay only into two S-wave mesons$^{\left[ 20\right] }$ and
equation $\left( 18\right) $ shows that a GE-hybrid does decay only into a
channel with one S-wave meson and one P-wave meson.

The same selection rules have been advocated for the light hybrid mesons$^{%
\left[ 19\right] }\ .$

In ref.$\left[ 17\right] $,the selection rule for GE-hybrid is proved in any
potential model.

\subsection{The decay in two ground state mesons}

The hybrid at $\sim 4.3GeV$ can decay in $D^{0}\overline{D^{0}}$, $%
D^{+}D^{-} $, $D_{s}^{+}D_{s}^{-}$, $D^{\ast 0}\overline{D^{0}}$, $D^{0}%
\overline{D^{\ast 0}}$, $D^{\ast 0}\overline{D^{\ast 0}}$, $D^{\ast
+}D^{\ast -}.$

Table 3 shows the results for the partial decay widths; the decays in $%
D^{\ast +}D^{\ast -}$and $D^{\ast 0}\overline{D^{\ast 0}}$ for $S=1$ are
suppressed by the spin overlap.

\begin{center}
\begin{equation*}
\underset{\text{\textit{Table 3: Decay widths of the (M=4.26) hybrid in
(S+S)-standard mesons (in }}\alpha _{s}\text{\textit{MeV).}}}{_{\underset{%
\text{\textit{.}}}{ 
\begin{tabular}{|c|c|c|c|}
\hline
$L$ & $0$ & $1$ & $2$ \\ \hline
$\Gamma _{D^{0}\bar{D}^{0}}$ & $129.5$ & $388.5$ & $647.5$ \\ \hline
$\Gamma _{D^{+}D^{-}}$ & $135.1$ & $406$ & $676.2$ \\ \hline
$\Gamma _{D_{s}^{+}D_{s}^{-}}$ & $142.8$ & $428.4$ & $714$ \\ \hline
$\Gamma _{D^{\ast 0}\overline{D^{0}}}=\Gamma _{D^{\ast 0}\overline{D^{\ast 0}%
}}$ & $0.00$ & $0.00$ & $0.00$ \\ \hline
$\Gamma _{D^{\ast +}D^{\ast -}=}\Gamma _{D^{\ast 0}\overline{D^{\ast 0}}}%
\text{\ } 
\begin{tabular}{l}
$S=0$ \\ 
$S=1$ \\ 
$S=2$%
\end{tabular}
$ & 
\begin{tabular}{l}
$30.8$ \\ 
\multicolumn{1}{c}{$0.00$} \\ 
$49.3$%
\end{tabular}
& 
\begin{tabular}{l}
$92.4$ \\ 
\multicolumn{1}{c}{$0.00$} \\ 
$369.6$%
\end{tabular}
& 
\begin{tabular}{l}
$1.47$ \\ 
\multicolumn{1}{c}{$0.00$} \\ 
$24.5$%
\end{tabular}
\\ \hline
$\Gamma _{tot}\left[ 1^{--}\left( 4.26\right) \right] $ & $982.1$ & $2463.3$
& $2914.1$ \\ \hline
\end{tabular}
}}}
\end{equation*}
\end{center}

The total decay width of a hybrid charmonium meson with mass $M=4.26GeV$ is
very large; such state does not emerge from the continuum of the two mesons
spectrum.

\subsection{The decay in (S-meson+P-meson)}

We give the results of the decay widths in Table 4.

\begin{center}
\begin{equation*}
\underset{\text{\textit{Table 4: Partial decay widths of the (M=4.3) hybrid
in (L+S)-standard mesons (in }}\alpha _{s}\text{\textit{MeV).}}}{ 
\begin{tabular}{|c|c|}
\hline
$\Gamma _{D_{1}(2420)\overline{D^{0}}}=\Gamma _{\overline{D_{1}}(2420)D^{0}}$
$\simeq \Gamma _{D_{1}^{+}(2420)D^{-}}=\Gamma _{D_{1}^{-}(2420)D^{+}}$ & $%
\frac{107}{4}$ \\ \hline
\end{tabular}
}
\end{equation*}
\end{center}

We find each decay rate of $\sim \frac{107}{4}MeV$, which is enough small
compared to the level spacing, to generate observable resonance (the total
decay width $\sim 107MeV$). It should be noted that the decay width of the $%
Y\left( 4260\right) $ is $\sim 90MeV$.

\subsection{Is the $Y\left( 4260\right) $ a mixed charmonium-hybrid meson?}

In order to study this possibility, we can refer to $\left[ 20\right] .$ The
calculated transition Hamiltonian between a $c\overline{c}g$ $1^{--}$ hybrid
and the conventional $c\overline{c}$ meson (such as $\Psi \left( 3S\right) $%
) gives very small amplitudes, then very small resulting angles, which
exclude the hypothesis to observe a mixed state .

\section{Results and Conclusion}

Therefore, according to the results, we may conclude that (in our model) a $%
J^{PC}=1^{--}$ charmonium hybrid meson should have a mass around $\sim
4.3GeV $ and decay preferably to $D_{1}\left( 2420\right) \overline{D}$. It
is important to check the existence of such state; more experimental
investigations are needed, searching for two-body decay channels (or $%
D^{\ast }D\pi $ channels), and confirmations on the mass and the decay width.

\begin{center}
$\FRAME{itbpFU}{1.2246in}{0.921in}{0in}{\Qcb{Fig1. Dominant decay process}}{%
}{j8k3to00.jpg}{\special{language "Scientific Word";type
"GRAPHIC";maintain-aspect-ratio TRUE;display "USEDEF";valid_file "F";width
1.2246in;height 0.921in;depth 0in;original-width 8.3333in;original-height
6.25in;cropleft "0";croptop "1";cropright "1";cropbottom "0";filename
'C:/Documents and
Settings/Administrateur/Bureau/J8K3TO00.JPG';file-properties "XNPEU";}}$

$\FRAME{itbpFU}{1.235in}{0.9288in}{0in}{\Qcb{Fig 2. OZI forbidden process}}{%
}{j8k3to01.jpg}{\special{language "Scientific Word";type
"GRAPHIC";maintain-aspect-ratio TRUE;display "USEDEF";valid_file "F";width
1.235in;height 0.9288in;depth 0in;original-width 8.3333in;original-height
6.25in;cropleft "0";croptop "1";cropright "1";cropbottom "0";filename
'C:/Documents and
Settings/Administrateur/Bureau/J8K3TO01.JPG';file-properties "XNPEU";}} $

$\FRAME{itbpFU}{1.2557in}{0.9444in}{0in}{\Qcb{Fig 3. Direct production of $%
\protect\pi ^{+}\protect\pi ^{-}J/\protect\psi $}}{}{j8k3tp02.jpg}{\special%
{language "Scientific Word";type "GRAPHIC";maintain-aspect-ratio
TRUE;display "USEDEF";valid_file "F";width 1.2557in;height 0.9444in;depth
0in;original-width 8.3333in;original-height 6.25in;cropleft "0";croptop
"1";cropright "1";cropbottom "0";filename 'C:/Documents and
Settings/Administrateur/Bureau/J8K3TP02.JPG';file-properties "XNPEU";}}$
\end{center}

\end{document}